\documentclass[12pt,titlepage]{article}
\usepackage[round,numbers,sort&compress]{natbib}
\usepackage{times}
\usepackage{graphics}% Include figure files
\usepackage{epsfig,calc}% Include figure files

\usepackage{bm}% bold math
\usepackage[usenames]{color}

\usepackage{amsmath,amssymb,chemarr}

\def\btt1{{\tt$\backslash$\string1}}%

\def\AmS{{\protect\the\textfont2
        A\kern-.1667em\lower.5ex\hbox{M}\kern-.125emS}}

\citeindextrue
\newcommand{\kt}{k_{\text{B}}T}
\newcommand{\eb}{\varepsilon_{\text{b}}}
\newcommand{\Np}{N_\text{p}}
\newcommand{\Nc}{N_\text{c}}
\newcommand{\epc}{\varepsilon_\text{pc}}
\newcommand{\ess}{\varepsilon_\text{ss}}
\newcommand{\epp}{\varepsilon_\text{pp}}
\newcommand{\dc}{n_\text{circ}}
\newcommand{\tf}{t_\text{f}}

\newcommand{\nnuc}{n_\text{nuc}}
\newcommand{\tnuc}{\tau_\text{nuc}}
\newcommand{\ov}{\hat{\mathbf{\Omega}}}
\newcommand{\ova}{\ov_\text{a}}
\newcommand{\rvec}{\mathbf{r}}
\newcommand{\toler}{\delta}

\newcommand{\nn}{\hat{\mathbf{n}}}
\newcommand{\dv}{\hat{\mathbf{r}}_{ij}}
\newcommand{\db}{\hat{\mathbf{d}}_\text{b}}

\newcommand{\conc}{c_0}

\newcommand{\cv}{c_\text{v}}
\newcommand{\crand}{c_\text{r}}
\begin{document}
%\baselineskip=1.3\baselineskip
%\title{Proposal Outline}
%\maketitle

\title{Mechanisms of capsid assembly around a polymer}
\author{Aleksandr Kivenson\\
%uncomment next 2lines for biophysj
Department of Physics\\
Brandeis University, Waltham, MA, 02454
		\and Michael~F.~Hagan \thanks{ Corresponding author. Address: hagan@brandeis.edu} \\
Department of Physics\\
Brandeis University, Waltham, MA, 02454
}

%comment for biophysj
%\affiliation{Department of Physics, Brandeis University, Waltham, MA, 02454}
%\author{Oren M. Elrad}
%\affiliation{Department of Physics, Brandeis University, Waltham, MA, 02454}

\date{\today}
% Running head
\pagestyle{myheadings}

%What is this?
\markright{Mechanisms of capsid assembly around a polymer}

% We are done with the headers, the actual document starts here
%\begin{document}

% generate the title page from the info in the headers above
\maketitle

\begin{abstract}
Capsids of many viruses assemble around nucleic acids or other polymers. Understanding how the properties of the packaged polymer affect the assembly process could promote biomedical efforts to prevent viral assembly or nanomaterials applications that exploit assembly. To this end, we simulate on a lattice the dynamical assembly of closed, hollow shells composed of several hundred to 1000 subunits, around a flexible polymer. We find that assembly is most efficient at an optimum polymer length that scales with the surface area of the capsid; significantly longer than optimal polymers often lead to partial-capsids with unpackaged polymer `tails' or a competition between multiple partial-capsids attached to a single polymer. These predictions can be tested with bulk experiments in which capsid proteins assemble around homopolymeric RNA or synthetic polyelectrolytes. We also find that the polymer can increase the net rate of subunit accretion to a growing capsid both by stabilizing the addition of new subunits and by enhancing the incoming flux of subunits; the effects of these processes may be distinguishable with experiments that monitor the assembly of individual capsids.
\end{abstract}


\clearpage
%\maketitle

%Do we need this section label?
%\section{Introduction}
The self-assembly of ordered structures is crucial in biology and is now providing a route to develop novel nanostructured materials. The success of assembly is governed by a competition between thermodynamics and kinetics, since metastable disordered states (kinetic traps) can impede the formation of an ordered thermodynamic ground state \cite{Whitesides2002,Ceres2002,Jack2007,Whitelam2009,Hagan2006,Licata2007,Rapaport2008}. Viral proteins are a paradigm for successful assembly -- during the replication of a typical virus, hundreds of proteins selectively form a protein shell, or capsid, that encases the viral nucleic acid. Single-stranded RNA virus capsids assemble around their RNA, and require RNA (or other polyanions \cite{Bancroft1969,Sikkema2007,Hu2008,Sun2007,Dixit2006}) to assemble at physiological conditions. How the packaged polymer promotes assembly is poorly understood because assembly intermediates are transient and thus challenging to characterize with experiments. Therefore, this paper examines a highly simplified model for capsid assembly around a flexible polymer, which yields experimentally testable predictions for the relationships of polymer length and solution conditions to assembly kinetics and assembly yields. Understanding how the polymer effects encapsulation dynamics could spur development of antiviral drugs that block assembly and provide critical knowledge to exploit capsids for use as drug delivery vehicles or gene therapy vectors.

In this first simulation study of the dynamics of polymer encapsidation we aim for the simplest possible description of capsid assembly around a polymer. Because successful assembly must avoid kinetic traps, we require a model with no \emph{a priori} assumptions about assembly pathways or the structures that emerge from assembly. The essential ingredients are: (1) the protein and polymer units are space filling, (2) the lowest energy state for the capsid is a hollow shell, and (3) there are short-ranged attractive interactions (representing screened electrostatics) between the polymer and protein units that favor encapsulation of the polymer. We arrive at a lattice model for protein and polymer units with pairwise attractions (Fig.~\ref{BondDiagram} and section Model), with assembly simulated with dynamic Monte Carlo. Because our questions are not specific to a particular capsid symmetry, we consider a cubic lattice and `capsids' for which a cubic shell is the ground state. The model is general, however, and could be implemented on a quasicrystalline lattice that allows icosahedral symmetry.

Elegant experiments have studied capsid assembly around ssRNA (e.g.\cite{Fox1994, Valegard1997, Johnson2004,Tihova2004,Krol1999,Stockley2007,Toropova2008,Johnson2004a}), but it is difficult to relate individual nucleic acid properties to assembly behavior because nucleic acid molecules with different sequences can have dramatically different secondary and tertiary structures \cite{Yoffe2008} and the fact that capsids assemble around synthetic polyelectrolytes  \cite{Bancroft1969,Sikkema2007,Hu2008} and nanoparticles  \cite{Sun2007,Dixit2006} demonstrates that properties specific to nucleic acids are not required for capsid formation or cargo packaging. Therefore in this work we primarily focus on experimental model systems in which capsid proteins assemble around synthetic polyelectrolytes \cite{Bancroft1969,Sikkema2007,Hu2008} or homopolymeric RNA. To begin to understand the effects of RNA-RNA base-pairing, we consider an extension to our model in which there are short-ranged attractive interactions between polymer segments. Although these attractions specifically represent a linear polymer in a poor solvent, they could shed light on assembly around RNA molecules, which form compact structures in solution due to base-pairing \cite{Yoffe2008}.

Under optimal conditions for assembly, we find that capsid growth and polymer incorporation proceed in concert, with the polymer forming a dense adsorbed layer on the partial capsid intermediate that stabilizes the addition of new capsid subunits. Whereas assembly is highly efficient under these conditions, longer than optimal polymers or strong interactions lead to several characteristic forms of kinetic traps. These should be identifiable in capsid assembly experiments through electron microscopy. The simulations demonstrate that assembly around a polymer can proceed by mechanisms unlike those identified for empty capsid formation, such as the diffusion of unassembled protein subunits along the polymer \cite{Hu2007b}.

Due to its simplicity and the lattice-implementation, our model provides significant computational speed up as compared to previously developed models, and allows simulation of time scales that correspond to minutes in real time and capsids with up to thousands of subunits. Thus, we are able to build upon prior modeling studies of capsid-polymer assembly that are equilibrium calculations and/or postulate particular assembly pathways and structures \cite{Devkota2009,Forrey2009,Harvey2009,Belyi2006,Angelescu2006,Schoot2005,Zhang2004a,Lee2008,Angelescu2008,Zandi2009,McPherson2005,Rudnick2005,Hu2007b} (for a nice review see \cite{Angelescu2008a}). Because our predictions for assembly yields and assembly rates are experimentally testable, we hope to motivate experiments that build upon prior experimental work investigating structures (e.g.\cite{Sorger1986,Fox1994, Valegard1997, Johnson2004,Tihova2004,Krol1999,Stockley2007,Toropova2008}) and kinetics\cite{Johnson2004a} of viral proteins assembling around nucleic acids.

\section*{Mechanisms of polymer encapsidation}
\label{tempLabel}
\begin{figure*}
\begin{center}
\includegraphics[width = 0.99\textwidth ]{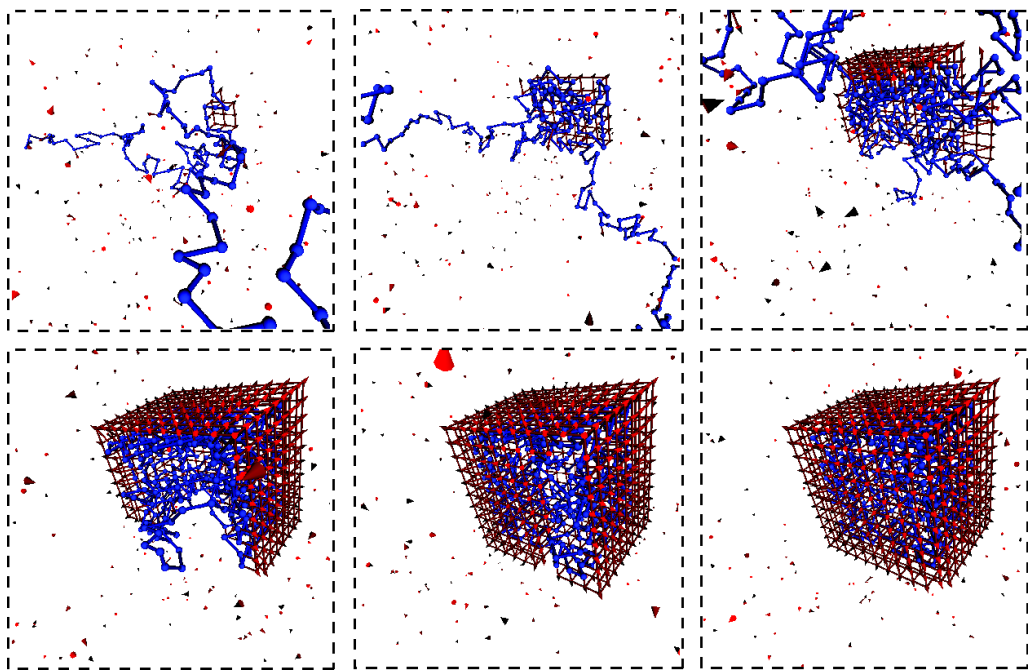}
\end{center}
\caption{Snapshots from a typical assembly trajectory; a small region of the simulation box is shown. Parameters are $\Np=350$, $\epc=5.75 \kt$, and $\Nc=488$.\label{fig:snapshots}}
\end{figure*}

\begin{figure*}
\begin{center}
\includegraphics[width = 0.99\textwidth ]{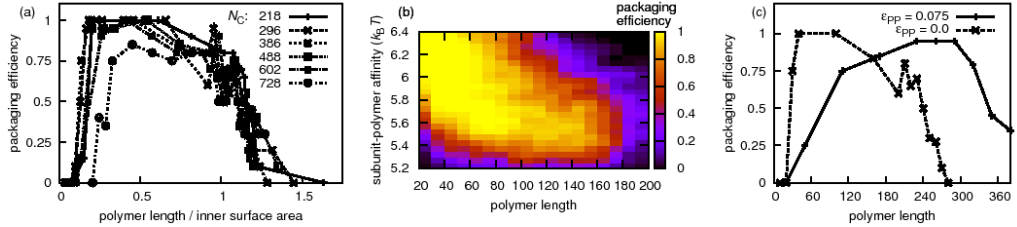}
\end{center}
\caption{Packaging efficiencies depend on polymer length, capsid size, polymer-subunit affinity, and time. {\bf (a)} Packaging efficiencies are shown as a function of the ratio of polymer length to inner surface area, for capsids with indicated capsid sizes $\Nc$ and $\epc=5.75 \kt$. {\bf (b)} Packaging efficiencies are shown as a function of polymer length $\Np$ and polymer-subunit affinity $\epc$  for capsids with size $\Nc=296$. Results are shown for $\tf=2\times10^8$ MC sweeps. {\bf (c)} Packaging efficiencies are compared with polymer-polymer attractions ($\epp=0.075 \kt$, $\epc=5.25 \kt$) and without ($\epp=0.0 \kt$, $\epc=5.75 \kt$). The lowest-energy capsid size is $\Nc=386$, but somewhat larger capsids can form around long polymers.\label{successFraction}}
\end{figure*}

To understand the influence of polymer properties on capsid assembly, we performed simulations for a range of binding energies $\eb$, capsid sizes $\Nc$, polymer-subunit interaction energies $\epc$ and polymer lengths $\Np$. The parameters $\eb$ and $\epc$ could be experimentally controlled by varying solution $p$H or ionic strength \cite{Ceres2002,Kegel2004}. We will discuss simulations with $\eb=5.85 \kt$ and capsid free subunit volume fraction $0.5 \%$; spontaneous assembly of empty capsids at this subunit concentration requires $\eb \gtrsim 6.5 \kt$. Except for Fig.~\ref{successFraction}c, all simulations consider no polymer-polymer attractions ($\epp=0.0 \kt$). For most parameter sets that lead to successful polymer encapsidation, assembly first requires nucleation of a small partial-capsid associated to the polymer, which is followed by a growth phase in which subunits reversibly bind to the partial-capsid (Fig.~\ref{fig:snapshots}). Finally, there can be a completion phase, during which addition of the final few subunits is delayed until the polymer is entirely incorporated within the capsid.

\begin{figure}
\includegraphics[width = 0.49\columnwidth ]{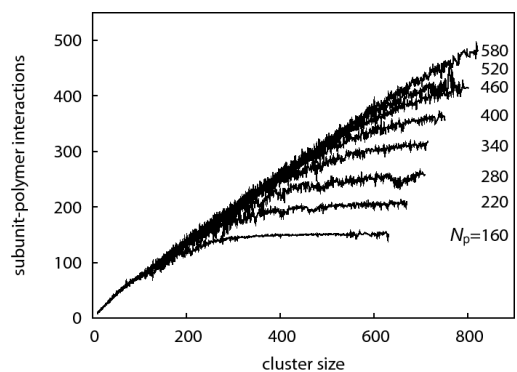}
\caption{The average number of favorable polymer-capsid contacts is shown as a function of partial-capsid intermediate size for assembly trajectories of capsids with size $\Nc=728$, for various indicated polymer lengths. The polymer-subunit affinity is $\epc=5.75 \kt$.\label{histogram}}
\end{figure}

Throughout the growth phase, the polymer adsorbs onto the capsid intermediate in a dense layer with relatively short loops and one or two long tails. The nature of the adsorbed layer is independent of polymer length and partial-capsid size until the entire polymer is adsorbed, in the sense that the number of polymer-capsid interactions (per subunit in the partial-capsid) is independent of polymer length until the entire polymer is adsorbed, as shown in Fig. \ref{histogram} for $\epc=5.75 \kt$ (see also Fig. S11 in the Supporting Material).  This independence is observed for all $\epc$ for which assembly occurred, although the ratio of polymer-capsid contacts to partial-capsid size decreases with decreasing $\epc$. The formation of a dense layer with a small number of long loops or tails for a polymer adsorbing onto a small surface was predicted theoretically \cite{Aubouy1998}.

\begin{figure}
\begin{center}
\includegraphics[width = 0.49\columnwidth ]{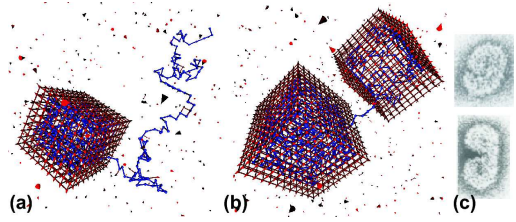}
\end{center}
\caption{Examples of configurations that rarely lead to complete polymer incorporation. {\bf (a)} Capsid closure progressed faster than complete polymer incorporation, trapping an exposed polymer tail, which blocks insertion of the final subunit. {\bf (b)} Two partial-capsids nucleated on the same polymer and grew to nearly complete capsids. Parameters for both cases are $\Nc=488$, $\Np=400$, and $\epc=5.75 \kt$.  {\bf (c)} Malformed polymer-capsid assemblies observed experimentally (figure adapted from Ref.~\cite{Sorger1986}).\label{traps}}
\end{figure}

{\bf Trapped configurations limit the length of polymer that can be spontaneously encapsidated.}
The fact that the amount of polymer incorporated within a partial-capsid is independent of polymer length constrains the length of polymer that can be efficiently packaged during assembly. In particular, assembly around long polymers frequently results in configurations such as shown in Fig.~\ref{traps}a, in which capsid closure was faster than polymer incorporation, trapping a polymer tail outside of the capsid. Complete polymer incorporation and assembly of the final few capsid subunits requires the polymer tail to retrace its contour into the capsid. However, few additional polymer-capsid interactions result during polymer retraction and additional capsid subunit-subunit interactions are only possible after the entire tail is inside the capsid. Furthermore, as shown by de Gennes \cite{Gennes1975}, the time for a polymer tail to retrace its contour is exponential in the length of the tail.  Hence, there can be a long completion phase in which assembly is stalled until the polymer tail completely retracts; polymer incorporation and capsid completion are rarely observed in our simulations after a capsid entraps a long polymer tail. Complete incorporation becomes even thermodynamically unfavorable above a certain polymer length, as first suggested by the equilibrium arguments of van der Schoot and Bruinsma \cite{Schoot2005}.

%\begin{figure}
%%\begin{center}
% \includegraphics[width = 0.9\columnwidth ]{./hist-labelled.eps}
%%\end{center}
%\caption{\label{histogram} The average number of favorable polymer-capsid contacts is shown as a function of partial-capsid intermediate size for assembly trajectories of capsids with size $\Nc=728$, for various indicated polymer lengths. The polymer-subunit affinity is $\epc=5.75 \kt$.
%}
%\end{figure}

An example of a second class of configurations that impede complete polymer incorporation is shown in Fig. \ref{traps}b. This `dumbbell' configuration is the usual outcome if two capsids nucleate on the same polymer and grow to significant size before coming into proximity; geometries of large partial-capsids are rarely compatible enough for a successful fusion event without significant subunit dissociation. Hence, dumbbell configurations are common for parameter sets for which capsid nucleation rates are significantly larger than capsid growth rates (see below). Completion of a capsid from this configuration is unlikely, since it would require complete retraction of the polymer from one of the capsids, which has a high free energy barrier (and is not thermodynamically favorable for polymers beyond a certain length).

Configurations with pinched polymer tails (Fig. \ref{traps}a) usually lead to dumbbell configurations if the time to nucleate a new capsid, which is inversely proportional to the length of the tail (see below) is shorter than that tail retraction time (exponential in tail length). We note that the dumbbell capsid configuration resembles malformed capsids that have been observed in experiments (e.g. Fig. \ref{traps}c)\cite{Sorger1986}. Thus, as discussed below, our prediction that configurations similar to those shown in Fig.~\ref{traps} will increase in frequency as the polymer length increases can be tested with imaging experiments.

{\bf Polymer encapsidation efficiency is nonmonotonic with respect to polymer length and polymer-subunit interaction strength.}
An experimentally accessible measure of assembly effectiveness is the packaging efficiency, or the fraction of trajectories in which a polymer is incorporated inside a complete capsid. Simulated packaging efficiencies are shown as a function of polymer length for varying capsid sizes (Fig. \ref{successFraction}a) and time (Fig. S9), with a complete capsid defined as a hollow shell with no gaps. For all times and capsid sizes, there is an optimal polymer length for which efficiency is maximal. Polymer lengths are normalized by the inner capsid surface area in Fig.~\ref{successFraction}a to show that the optimal polymer length is proportional to the number of polymer-subunit contacts in a complete capsid, which is proportional to the capsid size $N$. Note that the polymer radius of gyration, (see Fig. S8), is as much as thirty times the radius of the capsid for the longest polymers, consistent with the experimental observation that polystyrene sulfonate molecules with radii of gyration much larger than capsid size can be incorporated in cowpea chlorotic mottle virus capsids \cite{Bancroft1969,Hu2008}.

The fact that encapsidation efficiency is proportional to the driving force of polymer-subunit attractive interactions rather than capsid volume is consistent with theoretical equilibrium studies for polyelectrolyte encapsidation\cite{Schoot2005,Belyi2006,Siber2008}, noting however that polymer segment-segment electrostatic repulsions are accounted for only by segment excluded volume in our model. This observation can be understood by noting that the interactions in our model (and electrostatic interactions at physiological conditions) are short ranged; hence polymer-capsid interactions are confined to the layer at the inner capsid surface.  However, both thermodynamics and kinetics play an important role in limiting packaging efficiencies in our simulations. Below the optimal length, increasing polymer length provides a stronger thermodynamic driving force for assembly and enables faster nucleation and growth rates, as discussed below. Larger than optimal polymers also drive rapid capsid growth, but they tend to engender traps (discussed above) that block assembly. Similarly, the thermodynamically optimal length increases with polymer-subunit interaction strength $\epc$, but the optimal polymer length measured in our dynamical simulations decreases with increasing $\epc$ (Fig.~\ref{successFraction}b) because stronger interactions increase the propensity for kinetic traps. The interaction strength $\epc$ could be experimentally controlled by varying ionic strength.

\begin{figure*}
\begin{center}
\includegraphics[width = 0.99\textwidth ]{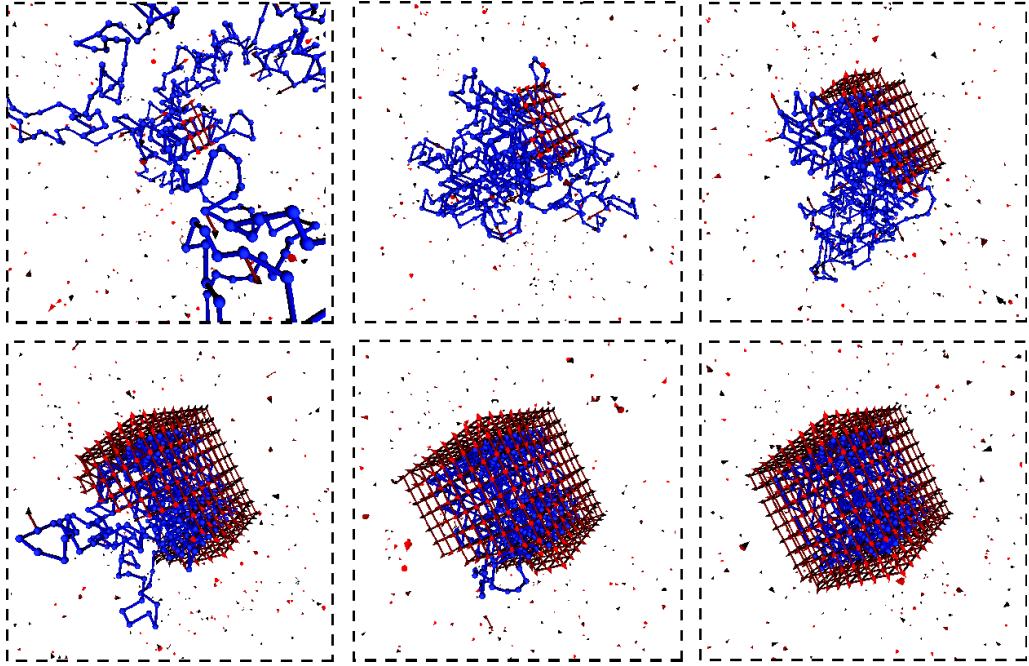}
\end{center}
\caption{Snapshots from a typical assembly trajectory with polymer-polymer attractions; a small region of the simulation box is shown. Parameters are $\Np=410$, $\epc=5.25 \kt$, $\epp=0.075 \kt$, and lowest energy capsid size $\Nc=386$.\label{fig:poorSolventTrajectory}}
\end{figure*}

{\bf Polymer-polymer attractions enable packaging of longer polymers.} Packaging efficiencies are shown as a function of polymer length for a polymer with segment-segment attractions ($\epp=0.075 \kt$) in Fig.~\ref{successFraction}c, where we see that polymers well above the length threshold for the case without attractions are packaged with nearly 100\% efficiency. This dramatic improvement in efficiency occurs because interaction with the assembling capsid causes the polymer to collapse into compact configurations (Fig.~\ref{fig:poorSolventTrajectory}), reducing the likelihood of incomplete polymer incorporation. This effect is even more significant for stronger attractions (past the coil-globule transition). We note that polymer attractions increase nucleation rates by decreasing the free energy of the critical nucleus (see the next section) and thus can increase the frequency of double nucleation. For this reason the results in Fig.~\ref{successFraction}c are shown for a decreased value of $\epc=5.25$. Furthermore, because the volume of an assembled capsid may vary somewhat, very long polymers are sometimes encapsidated because they drive the formation of a capsid larger than the optimal size dictated by subunit-subunit interactions.  This effect is much more prominent here than in simulations without polymer segment-segment attractions.

\section*{Nucleation and growth rates}
To understand the effect of system parameters on overall assembly rates, we measured durations of each phase of assembly (nucleation, growth, completion) as functions of polymer length and polymer-subunit interaction strength. For each trajectory, we define the nucleation time $\tnuc$ as the last timepoint for which the largest cluster was smaller than the critical nucleus size ($\nnuc=8$ subunits, see Fig. S10).  The growth time then corresponds to the interval between nucleation and containment, where a polymer is `contained' when within an assemblage of subunits that does not permit passage of 2x2x2 or larger cube; this definition distinguishes the growth phase from the completion phase described above.

\begin{figure*}
\includegraphics[width = 0.99\textwidth ]{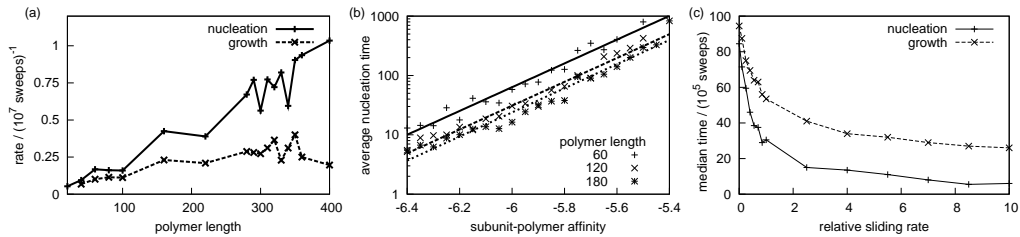}
\caption{{\bf (a)} Average rates for the nucleation and growth phases of assembly are shown as functions of polymer length, with $\Nc=488$ and $\epc=5.75 \kt$. {\bf (b)} Nucleation times decrease exponentially as subunit-polymer affinity increases.  {\bf (c)} Capsid growth rates depend on the subunit sliding rate. The median nucleation and growth times are shown for simulations with sliding moves (defined in the text) attempted with frequencies of 0, 1, and 10 (relative to attempt frequencies for ordinary subunit motions). Parameters are $\Np=100$, $\epc=5.75 \kt$, and $\Nc=296$.\label{growthTimes}}
\end{figure*}

{\bf Nucleation and growth rates increase with polymer length and polymer-subunit interaction strength.}
As shown in Fig.~\ref{growthTimes}, nucleation rates ($\tnuc^{-1}$) increase linearly with polymer length and exponentially with polymer-subunit interaction strength: $\tnuc^{-1} \sim \Np\exp[-\alpha(\nnuc-1) \epc/\kt]$.  %the rest of this paragraph could be moved to the SI if necessary XXX
These dependencies can be understood by noting that spontaneous nucleation of empty capsids does not occur for these system parameters, so nucleation requires that small partial-capsid intermediates are stabilized through interactions with the polymer.  Consistent with modeling studies of empty capsids  \cite{Endres2002,Hagan2009b} and assembly on nanoparticles  \cite{Hagan2008,Hagan2009},  the system rapidly builds up pre-nucleation intermediates with Boltzmann-weighted concentrations  $c_n\simeq c_0 \Np \exp[-(G_n + \alpha n\epc)/\kt]$ with $n$ the intermediate size, $G_n$ the subunit-subunit interaction free energy of intermediate $n$, and  the number of adsorbed intermediates is proportional to the polymer length $\Np$ for a fixed polymer concentration.  The parameter $\alpha$ is the number of polymer-subunit interactions per capsid subunit. The nucleation rate can then be expressed as \cite{Hagan2008}  $\tnuc^{-1}\simeq  c_0 c_{n_\text{nuc}-1}$ (see the SI for further analysis).

Growth rates also increase with increasing polymer length and polymer-subunit interaction strength, but saturate at a limiting value (Fig. \ref{growthTimes}a).  This trend reflects two mechanisms by which the polymer can influence capsid growth. First, binding to the polymer stabilizes partial-capsid intermediates; this is a thermodynamic effect that increases the net rate of assembly by decreasing the rate of subunit desorption from adsorbed intermediates. The net stabilization is proportional to the number of subunit-polymer contacts in a partial-capsid intermediate, which is independent of polymer length until the polymer is completely incorporated (Fig. \ref{histogram}).   Therefore, capsid growth times are significantly longer for polymers that are completely incorporated before about two thirds of the capsid has assembled, but depend only weakly on polymer length for longer polymers. Similarly, the effect of increasing  $\epc$ saturates when the unbinding rates of polymer-stabilized subunits become small compared to association rates.

{\bf Subunit sliding.}
The second effect of the polymer on growth times is purely kinetic; as proposed by Hu and Shklovskii \cite{Hu2007b} the polymer can enhance the flux of subunits to binding sites on partial-capsid intermediates, through sliding or correlated diffusion of subunits along the polymer.  To characterize the role of sliding in our simulations, we performed simulations with extra `sliding' move attempts, in which a subunit interacting with a polymer is moved forward or backward by one polymer segment, and a subunit orientation is chosen at random from the set of orientations that enable a subunit-polymer attraction; the move is accepted or rejected according to the Metropolis criteria (full details in the SI). The subunit sliding rate (one-dimensional diffusion constant) was varied by changing the frequency of sliding moves relative to ordinary subunit motions; as shown in Fig.~\ref{growthTimes}c, nucleation and growth rates increase with the relative sliding rate. This observation indicates that subunit sliding enhances assembly rates, but shows that sliding affects nucleation and growth rates differently; the effect on nucleation will dominate under typical conditions, for which nucleation is rate limiting.

We note that another mechanism by which a polymer could enhance the flux of subunits to a partial capsid is through correlated polymer-subunit motions (i.e. the polymer reels an attached subunit in like a hooked fish). Previous works \cite{McPherson2005,Hagan2008,Harvey2009} have suggested that in the limit of high polymer-subunit affinities subunits adsorb on a polymer en masse and then collectively reorient to form a capsid. As discussed in Model our Monte Carlo simulations do not effectively explore this mechanism because they use single particle moves \cite{Whitelam2007,Whitelam2009}. We are currently exploring the importance of collective moves using off-lattice Brownian dynamics simulations of capsid assembly around the polymer [unpublished]. We believe that the use of single particle moves does not qualitatively affect the results for most parameter sets because the dependencies on system parameters and failure modes reported here are similar to those obtained with the Brownian dynamics simulations. The results suggest that collective moves are significant at high polymer-subunit interactions and/or subunit concentrations; however, the conclusion that overly strong interactions or high subunit concentrations lead to kinetic traps remains valid.

{\bf Viral genome specificity.} The observation that assembly rates will be nucleation-limited under conditions of optimal for assembly and the preceding analysis offers one possible mechanism by which viruses can preferentially package their genomic RNA over random RNA (e.g. \cite{D'Souza2005,Jouvenet2009}). In many viruses, the capsid protein has been shown to specifically and strongly bind to a short `packaging sequence' on the viral genome. Denoting the sequence-specific binding free energy as $\ess$, the ratio of assembly rates around viral and random RNA is given by $\tnuc^{-1}(\text{viral})/\tnuc^{-1}(\text{random})=\exp[-n_\text{s}(\ess-\epc)/\kt] \cv/\crand$, with $\cv$ and $\crand$ the concentrations of viral and random RNA molecules in the vicinity of the assembling capsid proteins and $n_\text{s}$ the length of the packaging sequence. Nearly complete specificity for the viral genome will occur for $n_\text{s}(\ess-\epc) \gg \ln \cv/\crand$.  We are exploring this prediction in simulations that explicitly represent packaging sequences, and note that recent observations that viral genomes have smaller sizes in solution than other RNA molecules could also play a role.

\section*{Implications for experiments}
{\bf Measuring polymer incorporation efficiencies in experiments. }
The simulations described in this work predict that capsids can assemble around flexible polymers with high yields, but that packaging efficiencies are nonmonotonic with respect to polymer length.
These predictions could be tested by measuring packaging efficiencies in experiments in which capsid proteins from viruses with single-stranded genomes assemble around synthetic polyelectrolytes \cite{Sikkema2007,Hu2008} or nucleic acids that do not undergo base-pairing, such as homopolymeric RNA. Our results suggest that packaging efficiencies will decrease and incompletely incorporated polymers or dumbbell capsids will appear as the polymer length is increased significantly beyond the viral genome length.

Experimentally measuring packaging efficiencies will require distinguishing well-formed capsids from failed assemblages (those that have unincorporated polymeric tails or are otherwise incomplete). Failed assemblages could be identifiable by TEM or AFM \cite{Sun2007}.  In the case of assembly around homopolymeric RNA, incompletely incorporated RNA can be identified by treating the capsid solution with RNase to digest unincorporated polymer, washing out RNase, disassembling capsids, and running the remaining polymer on a gel \cite{Grayson2006}. %We note that although the incorporation of long polystyrene sulfonate molecules within cowpea chlorotic mottle virus capsids was reported in Ref. \cite{Hu2008}, packaging efficiencies were not measured in that experiment.

The simulation results for polymer-polymer attractions (Fig.~\ref{successFraction}c) suggest that unincorporated polymer tails will be less prevalent in the case of polymers that form compact structures. Although this model is not meant to represent RNA, base-pairing of viral RNA molecules leads to compact structures in solution\cite{Yoffe2008}, and thus these results suggest that unincorporated polymer tails and dumbbell capsids will be less prevalent for RNA. It would therefore be interesting to compare the polymer length dependence of packaging experiments involving homopolymeric RNA with those involving viral RNA. It would be important to consider various sequences of viral RNA to distinguish the effects of base-pairing from effects of recognition sequences.

{\bf Measuring capsid growth rates in experiments.} The simulations demonstrate that overall capsid formation time distributions are a convolution of the time distributions for each of the three phases: nucleation, growth, and completion. Since the completion phase is likely to be difficult to monitor in experiments, measured assembly time distributions will include nucleation times and growth times. Incorporation efficiencies are highest in our simulations when nucleation of multiple capsids on a single polymer is unlikely, which requires that nucleation times are longer than or comparable to growth times. It may therefore be difficult to extract dependencies of capsid growth times on system parameters from bulk measurements (see Refs.~\cite{Hagan2009b,Ceres2002} for a discussion of this constraint for empty capsids). With experiments that monitor the assembly of individual capsids (e.g. Ref. \cite{Jouvenet2008}), however, it is possible to separate nucleation and growth phases as we have done for the simulations in this work; comparison of results from these experiments to simulated growth times could elucidate mechanisms of capsid growth after nucleation. Although single capsid assembly experiments have thus far relied upon confocal microscopy to visualize assembly of capsids on cell membranes, it might be possible to visualize the assembly of non-membrane-associated capsids using confocal microscopy or total internal reflectance microscopy by tethering RNA molecules to a surface.

\section*{Model}

\begin{figure}
\begin{center}
\includegraphics[width = 0.49\columnwidth ]{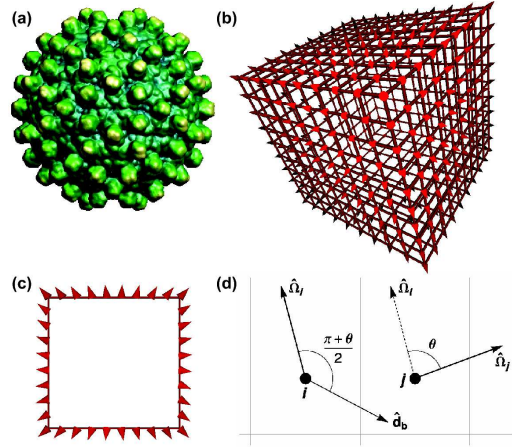}
\end{center}
\caption{The capsid model. {\bf (a)} The spike proteins of the hepatitis B virus (HBV) indicate the orientations of its subunits (image from Viper \cite{Natarajan2005}). {\bf (b)} A model capsid with $\Nc=488$ subunits (circumference $\dc=36$ subunits). Subunits are drawn as cones to indicate their orientations and lines are drawn between interacting subunits. {\bf (c)} A cross-section view of the model capsid. The spatial variation of subunit orientations in {\bf (b)} and {\bf (c)} can be compared to that of the spike proteins in HBV. {\bf (d)} The interaction geometry is shown for two model capsid subunits, $i$ and $j$. In this illustration, the orientation vectors $\ov_i$ and $\ov_j$ are in the plane of the figure and thus the rotation axis $\ov_a$ is perpendicular to that plane. The orientation of the bond vector $\db$ is determined by the angle $\theta$ between the two orientation vectors as described in the text. A favorable interaction for this configuration requires that $\theta$ satisfies Eq.~\ref{eq:ov} and that $\db$ and $\rvec_{ij}$ satisfy Eq.~\ref{eq:db}.\label{BondDiagram}}
\end{figure}

\noindent{\bf Capsid subunits.}  We enable simulation of large capsids and long time scales by representing capsid protein subunits as rigid bodies with discrete positions on a lattice and continuous orientations. Subunits have pairwise interactions comprised of excluded volume, represented by allowing a maximum of one subunit per lattice site, and attractions constructed such that the lowest energy states in the model are closed shells or `capsids' with a preferred number of subunits $\Nc$. As shown in  Fig.~\ref{BondDiagram}, the variations of subunit orientations within a model capsid can be compared with those in an actual icosahedral capsid. For a closed shell with a preferred size to be the lowest energy state, subunit-subunit interactions must (1) lead to a preferred large-scale curvature and (2) drive the formation of two dimensional manifolds rather than bulk structures. As has been the case for off-lattice models of capsid assembly \cite{Hagan2006,Hicks2006,Nguyen2007,Wilber2007,Nguyen2009}, these requirements are satisfied in our lattice model through interactions that simultaneously depend on relative orientations and positions of subunits (Eqs.~\ref{eq:db} and \ref{eq:ov}).
The model is designed to eliminate the influence of lattice structure on interaction free energies to the greatest extent possible.

To explain the interaction potential, we consider two neighboring subunits $i$ and $j$ with respective lattice positions $\rvec_i$ and $\rvec_j$, and unit orientation vectors $\ov_i$ and $\ov_j$ which have a relative angle $\theta=\cos^{-1}(\ov_i \cdot \ov_j)$, as shown in Fig.~\ref{BondDiagram}.  The subunits  experience a favorable interaction with energy $-\eb$ when two conditions (Eqs.~\ref{eq:ov} and \ref{eq:db}) are satisfied. First, an attraction requires that the subunit orientations are nearly consistent with the preferred circumference $\dc$ of the capsid
\begin{align}
|\theta - 2\pi/\dc| \le \toler
%\db \cdot \did \le \cos(\toler) %\;\; \mbox{or} \;\; \db^{-} \cdot \did^{-} \le \cos(\toler)
%2/\dc-\toler \le \phi \le 2/\dc+\toler
\label{eq:ov}
\end{align}
with $\toler$ the orientational specificity parameter. For all simulations in this work $\toler=\pi/30$.

The second requirement drives formation of a single layer shell. We define a bond vector $\db$ for the $i$, $j$ interaction, which is obtained by rotating $\ov_i$ by the angle $(\pi+\theta)/2$ around the rotation axis  $\ova = (\ov_i \times \ov_j)/|\ov_i \times \ov_j|$.  An attractive interaction requires that the displacement vector $\dv=\rvec_j-\rvec_i$ is the neighbor displacement vector that most closely parallels the bond vector $\db$:
\begin{align}
\rvec_j-\rvec_i  =  (\arg\,\max_{\nn} [\nn \cdot \db]) %  \;\; \mbox{or} \;\; \dv =( \arg\,\max_{\nn} [\nn \cdot \db^{-}])
%\db \cdot \did   \le  \cos(\toler)
\label{eq:db}
\end{align}
with $\{\nn\}$ the set of 26 neighbor lattice displacement vectors.

As shown in Fig.~\ref{BondDiagram}b, the interaction potential drives subunits to assemble into flat single-layer sheets with orientations that gradually rotate, which enables representation of curved structures even on a cubic lattice. The requirement Eq.~\ref{eq:db} drives the sheet to turn when subunit orientations reach a critical angle. Because of the finite angle tolerance $\toler$, the turn is stochastic and model capsids are not always perfect cubes. Near this critical angle, a single subunit could satisfy  Eqs.~\ref{eq:ov} and \ref{eq:db} with two different subunits each in a different lattice site with slightly different orientations. To avoid this possibility, subunits have `exclusion zones' located on the two of the 26 neighbor sites closest to the forward and backward extensions of the orientation unit vector respectively (see SI Figs. S6, S7). A subunit position cannot overlap with an exclusion zone of another subunit, but multiple exclusion zones can share the same lattice site.
The interaction geometry ensures that a subunit dimer will have the same interaction free energy for any lattice position and dimer orientation.%, but in a capsid the 8 corner subunits have one fewer interaction partners than other subunits.

{\bf Polymer.} The polymer is represented with the bond fluctuation model \cite{Carmesin1988,Shaffer1994}, modified so that polymer segments occupy only a single lattice site and have allowed bond lengths of 1 and $\sqrt{2}$.  Configurations in which polymer bonds cross are rejected; full details are given in the SI. The polymer radius of gyration scales as $\Np^{3/5}$ as expected for good solvent conditions (Fig. S8).
Polymer segments are endowed with a unit orientation vector, and experience interactions with energy $-\epc$ when (1) a capsid subunit is located one lattice site forward in the direction of the polymer orientation vector and (2) the negative of the subunit orientation vector points toward the polymer segment. This feature represents the fact that positive charges located on the inner surface of a capsid interact with encapsidated polyelectrolytes. To enable polymer-subunit bonds, polymer segments may occupy an exclusion zone defined by extending a subunit's orientation backward. In simulations with polymer-polymer attractions, polymer subunits experience isotropic attractive interactions with magnitude $\epp$ to other polymer subunits within the 26 closest neighbor sites.  Otherwise, there are no interactions between polymer segments except for excluded volume.

{\bf Simulations.} The simulations use dynamic Monte Carlo (MC) moves in which a subunit or a polymer segment is displaced to a lattice site randomly chosen from the set of 26 nearest neighbor sites and the current position, and a new orientation vector is chosen randomly from the unit sphere. This procedure assumes that translation and orientation relaxation times are comparable.  Moves are accepted or rejected according to the Metropolis criteria \cite{Rosenbluth1955}. To efficiently represent a dilute solution of polymer in excess capsid subunits, we use periodic boundary conditions and coupled the system to a bath of subunits with concentration $\conc$ by performing grand canonical MC moves in which subunits are inserted or deleted \cite{Hagan2008}. To maintain realistic dynamics, insertions and deletions are performed only in the outermost lattice layer (defined relative to the middle polymer segment) with a frequency consistent with the diffusion limited rate \cite{Hagan2008,Elrad2008}. To maintain computational feasibility with extremely long polymers, some simulations have a box side-length that is shorter than the full extension of the polymer. The side length $L$ was chosen based the relationship between polymer length $\Np$ and the confinement free energy $F_\text{conf}$ of an un-encapsidated polymer $F_\text{conf}\cong \Np^{9/4} L^{-15/4}$ \cite{Sakaue2006}, to maintain $F_\text{conf} \le 4 \kt$ (which is insignificant compared to total binding energies and entropies) and $L \ge 23$. There were no observed instances of multiple polymer images interacting with an assembling capsid.

%Because the Monte Carlo move set includes only single particle moves, some cooperative motions could be suppressed \cite{Whitelam2007,Whitelam2009}, in particular those in which polymer and capsid subunits move together. We believe that this constraint does not qualitatively affect the results because the dependencies on system parameters and failure modes reported here are similar to results obtained with our off-lattice Brownian dynamics simulations of capsid assembly around a polymer [unpublished]. For further discussion, see the SI.

{\bf Potential model extensions.}
Experiments with cowpea chlorotic mottle virus \cite{Bancroft1969,Hu2008} show that the virus can assemble with different triangulation (T) numbers around polymers of different sizes. This possibility could be examined with our model by incorporating subunit `conformational changes' which modify the preferred circumference, following Ref.~\cite{Elrad2008}. An explicit representation of RNA base-pairing would enable predictions of the differences between packaging homopolymers and viral RNA,  and should include the dependence of molecular rigidity on base-pairing as well as recognition sequences.

\section*{Supplementary Material}
Supplementary material is available online at http://www.brandeis.edu/~kivenson/kivenson2010si.pdf.

%\bibliography{./database-12-12-08}

\begin{thebibliography}{59}
\expandafter\ifx\csname natexlab\endcsname\relax\def\natexlab#1{#1}\fi

\bibitem[{Whitesides and Grzybowski(2002)}]{Whitesides2002}
Whitesides, G.~M., and B.~Grzybowski. 2002.
\newblock Self-assembly at all scales.
\newblock \emph{Science} 295:2418--2421.

\bibitem[{Ceres and Zlotnick(2002)}]{Ceres2002}
Ceres, P., and A.~Zlotnick. 2002.
\newblock Weak protein-protein interactions are sufficient to drive assembly of
  hepatitis b virus capsids.
\newblock \emph{Biochemistry} 41:11525--11531.

\bibitem[{Jack et~al.(2007)Jack, Hagan, and Chandler}]{Jack2007}
Jack, R.~L., M.~F. Hagan, and D.~Chandler. 2007.
\newblock Fluctuation-dissipation ratios in the dynamics of self-assembly.
\newblock \emph{Phys. Rev. E} 76:021119.

\bibitem[{Whitelam et~al.(2009)Whitelam, Feng, Hagan, and
  Geissler}]{Whitelam2009}
Whitelam, S., E.~H. Feng, M.~F. Hagan, and P.~L. Geissler. 2009.
\newblock The role of collective motion in examples of coarsening and
  self-assembly.
\newblock \emph{Soft Matter} 5:1251--1262.

\bibitem[{Hagan and Chandler(2006)}]{Hagan2006}
Hagan, M.~F., and D.~Chandler. 2006.
\newblock Dynamic pathways for viral capsid assembly.
\newblock \emph{Biophys. J.} 91:42--54.

\bibitem[{Licata and Tkachenko(2007)}]{Licata2007}
Licata, N.~A., and A.~V. Tkachenko. 2007.
\newblock Colloids with key-lock interactions: Nonexponential relaxation,
  aging, and anomalous diffusion.
\newblock \emph{Physical Review E} 76:7.

\bibitem[{Rapaport(2008)}]{Rapaport2008}
Rapaport, D. 2008.
\newblock The role of reversibility in viral capsid growth: A paradigm for
  self-assembly.
\newblock \emph{Phys. Rev. Lett.} 101:186101.

\bibitem[{Bancroft et~al.(1969)Bancroft, Hiebert, and Bracker}]{Bancroft1969}
Bancroft, J.~B., E.~Hiebert, and C.~E. Bracker. 1969.
\newblock Effects of various polyanions on shell formation of some spherical
  viruses.
\newblock \emph{Virology} 39:924--\&.

\bibitem[{Sikkema et~al.(2007)Sikkema, Comellas-Aragones, Fokkink, Verduin,
  Cornelissen, and Nolte}]{Sikkema2007}
Sikkema, F.~D., M.~Comellas-Aragones, R.~G. Fokkink, B.~J.~M. Verduin,
  J.~Cornelissen, and R.~J.~M. Nolte. 2007.
\newblock Monodisperse polymer-virus hybrid nanoparticles.
\newblock \emph{Org. Biomol. Chem.} 5:54--57.

\bibitem[{Hu et~al.(2008)Hu, Zandi, Anavitarte, Knobler, and Gelbart}]{Hu2008}
Hu, Y., R.~Zandi, A.~Anavitarte, C.~M. Knobler, and W.~M. Gelbart. 2008.
\newblock Packaging of a polymer by a viral capsid: The interplay between
  polymer length and capsid size.
\newblock \emph{Biophys. J.} 94:1428--1436.

\bibitem[{Sun et~al.(2007)Sun, DuFort, Daniel, Murali, Chen, Gopinath, Stein,
  De, Rotello, Holzenburg, Kao, and Dragnea}]{Sun2007}
Sun, J., C.~DuFort, M.~C. Daniel, A.~Murali, C.~Chen, K.~Gopinath, B.~Stein,
  M.~De, V.~M. Rotello, A.~Holzenburg, C.~C. Kao, and B.~Dragnea. 2007.
\newblock Core-controlled polymorphism in virus-like particles.
\newblock \emph{Proc. Natl. Acad. Sci. U. S. A.} 104:1354--1359.

\bibitem[{Dixit et~al.(2006)Dixit, Goicochea, Daniel, Murali, Bronstein, De,
  Stein, Rotello, Kao, and Dragnea}]{Dixit2006}
Dixit, S.~K., N.~L. Goicochea, M.~C. Daniel, A.~Murali, L.~Bronstein, M.~De,
  B.~Stein, V.~M. Rotello, C.~C. Kao, and B.~Dragnea. 2006.
\newblock Quantum dot encapsulation in viral capsids.
\newblock \emph{Nano Lett.} 6:1993--1999.

\bibitem[{Fox et~al.(1994)Fox, Johnson, and Young}]{Fox1994}
Fox, J.~M., J.~E. Johnson, and M.~J. Young. 1994.
\newblock Rna/protein interactions in icosahedral virus assembly.
\newblock \emph{Seminars in Virology} 5:51--60.

\bibitem[{Valegard et~al.(1997)Valegard, Murray, Stonehouse, vandenWorm,
  Stockley, and Liljas}]{Valegard1997}
Valegard, K., J.~B. Murray, N.~J. Stonehouse, S.~vandenWorm, P.~G. Stockley,
  and L.~Liljas. 1997.
\newblock The three-dimensional structures of two complexes between recombinant
  ms2 capsids and rna operator fragments reveal sequence-specific protein-rna
  interactions.
\newblock \emph{J. Mol. Biol.} 270:724--738.

\bibitem[{Johnson et~al.(2004{\natexlab{a}})Johnson, Tang, Johnson, and
  Ball}]{Johnson2004}
Johnson, K.~N., L.~Tang, J.~E. Johnson, and L.~A. Ball. 2004{\natexlab{a}}.
\newblock Heterologous rna encapsidated in pariacoto virus-like particles forms
  a dodecahedral cage similar to genomic rna in wild-type virions.
\newblock \emph{J. Virol.} 78:11371--11378.

\bibitem[{Tihova et~al.(2004)Tihova, Dryden, Le, Harvey, Johnson, Yeager, and
  Schneemann}]{Tihova2004}
Tihova, M., K.~A. Dryden, T.~V.~L. Le, S.~C. Harvey, J.~E. Johnson, M.~Yeager,
  and A.~Schneemann. 2004.
\newblock Nodavirus coat protein imposes dodecahedral rna structure independent
  of nucleotide sequence and length.
\newblock \emph{J. Virol.} 78:2897--2905.

\bibitem[{Krol et~al.(1999)Krol, Olson, Tate, Johnson, Baker, and
  Ahlquist}]{Krol1999}
Krol, M.~A., N.~H. Olson, J.~Tate, J.~E. Johnson, T.~S. Baker, and P.~Ahlquist.
  1999.
\newblock Rna-controlled polymorphism in the in vivo assembly of 180-subunit
  and 120-subunit virions from a single capsid protein.
\newblock \emph{Proc. Natl. Acad. Sci. U. S. A.} 96:13650--13655.

\bibitem[{Stockley et~al.(2007)Stockley, Rolfsson, Thompson, Basnak, Francese,
  Stonehouse, Homans, and Ashcroft}]{Stockley2007}
Stockley, P.~G., O.~Rolfsson, G.~S. Thompson, G.~Basnak, S.~Francese, N.~J.
  Stonehouse, S.~W. Homans, and A.~E. Ashcroft. 2007.
\newblock A simple, rna-mediated allosteric switch controls the pathway to
  formation of a t=3 viral capsid.
\newblock \emph{J. Mol. BIo.} 369:541--552.

\bibitem[{Toropova et~al.(2008)Toropova, Basnak, Twarock, Stockley, and
  Ranson}]{Toropova2008}
Toropova, K., G.~Basnak, R.~Twarock, P.~G. Stockley, and N.~A. Ranson. 2008.
\newblock The three-dimensional structure of genomic rna in bacteriophage ms2:
  Implications for assembly.
\newblock \emph{J. Mol. Biol.} 375:824--836.

\bibitem[{Johnson et~al.(2004{\natexlab{b}})Johnson, Willits, Young, and
  Zlotnick}]{Johnson2004a}
Johnson, J.~M., D.~A. Willits, M.~J. Young, and A.~Zlotnick.
  2004{\natexlab{b}}.
\newblock Interaction with capsid protein alters rna structure and the pathway
  for in vitro assembly of cowpea chlorotic mottle virus.
\newblock \emph{J. Mol. Biol.} 335:455--464.

\bibitem[{Yoffe et~al.(2008)Yoffe, Prinsen, Gopal, Knobler, Gelbart, and
  Ben-Shaul}]{Yoffe2008}
Yoffe, A.~M., P.~Prinsen, A.~Gopal, C.~M. Knobler, W.~M. Gelbart, and
  A.~Ben-Shaul. 2008.
\newblock Predicting the sizes of large rna molecules.
\newblock \emph{Proc. Natl. Acad. Sci. U. S. A.} 105:16153--16158.

\bibitem[{Hu and Shklovskii(2007)}]{Hu2007b}
Hu, T., and B.~I. Shklovskii. 2007.
\newblock Kinetics of viral self-assembly: Role of the single-stranded rna
  antenna.
\newblock \emph{Phys. Rev. E} 75:051901.

\bibitem[{Devkota et~al.(2009)Devkota, Petrov, Lemieux, Boz, Tang, Schneemann,
  Johnson, and Harvey}]{Devkota2009}
Devkota, B., A.~S. Petrov, S.~Lemieux, M.~B. Boz, L.~Tang, A.~Schneemann, J.~E.
  Johnson, and S.~C. Harvey. 2009.
\newblock Structural and electrostatic characterization of pariacoto virus:
  Implications for viral assembly.
\newblock \emph{Biopolymers} 91:530--538.

\bibitem[{Forrey and Muthukumar(2009)}]{Forrey2009}
Forrey, C., and M.~Muthukumar. 2009.
\newblock Electrostatics of capsid-induced viral rna organization.
\newblock \emph{J. Chem. Phys.} 131.

\bibitem[{Harvey et~al.(2009)Harvey, Petrov, Devkota, and Boz}]{Harvey2009}
Harvey, S.~C., A.~S. Petrov, B.~Devkota, and M.~B. Boz. 2009.
\newblock Viral assembly: a molecular modeling perspective.
\newblock \emph{Phys. Chem. Chem. Phys.} 11:10553--10564.

\bibitem[{Belyi and Muthukumar(2006)}]{Belyi2006}
Belyi, V.~A., and M.~Muthukumar. 2006.
\newblock Electrostatic origin of the genome packing in viruses.
\newblock \emph{Proc. Natl. Acad. Sci. U. S. A.} 103:17174--17178.

\bibitem[{Angelescu et~al.(2006)Angelescu, Bruinsma, and Linse}]{Angelescu2006}
Angelescu, D.~G., R.~Bruinsma, and P.~Linse. 2006.
\newblock Monte carlo simulations of polyelectrolytes inside viral capsids.
\newblock \emph{Phys. Rev. E} 73:041921.

\bibitem[{van~der Schoot and Bruinsma(2005)}]{Schoot2005}
van~der Schoot, P., and R.~Bruinsma. 2005.
\newblock Electrostatics and the assembly of an rna virus.
\newblock \emph{Phys. Rev. E} 71:061928.

\bibitem[{Zhang et~al.(2004)Zhang, Konecny, Baker, and McCammon}]{Zhang2004a}
Zhang, D.~Q., R.~Konecny, N.~A. Baker, and J.~A. McCammon. 2004.
\newblock Electrostatic interaction between rna and protein capsid in cowpea
  chlorotic mottle virus simulated by a coarse-grain rna model and a monte
  carlo approach.
\newblock \emph{Biopolymers} 75:325--337.

\bibitem[{Lee and Nguyen(2008)}]{Lee2008}
Lee, S., and T.~T. Nguyen. 2008.
\newblock Radial distribution of rna genomes packaged inside spherical viruses.
\newblock \emph{Phys. Rev. Lett.} 100.

\bibitem[{Angelescu et~al.(2008)Angelescu, Linse, Nguyen, and
  Bruinsma}]{Angelescu2008}
Angelescu, D.~G., P.~Linse, T.~T. Nguyen, and R.~F. Bruinsma. 2008.
\newblock Structural transitions of encapsidated polyelectrolytes.
\newblock \emph{European Physical Journal E} 25:323--334.

\bibitem[{Zandi and van~der Schoot(2009)}]{Zandi2009}
Zandi, R., and P.~van~der Schoot. 2009.
\newblock \emph{Biophys. J.} 96:9--20.

\bibitem[{McPherson(2005)}]{McPherson2005}
McPherson, A. 2005.
\newblock Micelle formation and crystallization as paradigms for virus
  assembly.
\newblock \emph{BioEssays} 27:447--458.

\bibitem[{Rudnick and Bruinsma(2005)}]{Rudnick2005}
Rudnick, J., and R.~Bruinsma. 2005.
\newblock Icosahedral packing of rna viral genomes.
\newblock \emph{Phys. Rev. Lett.} 94:038101.

\bibitem[{Angelescu and Linse(2008)}]{Angelescu2008a}
Angelescu, D.~G., and P.~Linse. 2008.
\newblock Viruses as supramolecular self-assemblies: modelling of capsid
  formation and genome packaging.
\newblock \emph{Soft Matter} 4:1981--1990.

\bibitem[{Sorger et~al.(1986)Sorger, Stockley, and Harrison}]{Sorger1986}
Sorger, P.~K., P.~G. Stockley, and S.~C. Harrison. 1986.
\newblock Structure and assembly of turnip crinkle virus .2. mechanism of
  reassembly invitro.
\newblock \emph{J. Mol. Biol.} 191:639--658.

\bibitem[{Kegel and van~der Schoot(2004)}]{Kegel2004}
Kegel, W.~K., and P.~van~der Schoot. 2004.
\newblock Competing hydrophobic and screened-coulomb interactions in hepatitis
  b virus capsid assembly.
\newblock \emph{Biophys. J.} 86:3905--3913.

\bibitem[{Aubouy and Raphael(1998)}]{Aubouy1998}
Aubouy, M., and E.~Raphael. 1998.
\newblock Scaling description of a colloidal particle clothed with polymers.
\newblock \emph{Macromolecules} 31:4357--4363.

\bibitem[{de~Gennes(1975)}]{Gennes1975}
de~Gennes, P.~G. 1975.
\newblock Reptation of stars.
\newblock \emph{Journal De Physique} 36:1199--1203.

\bibitem[{Siber and Podgornik(2008)}]{Siber2008}
Siber, A., and R.~Podgornik. 2008.
\newblock Nonspecific interactions in spontaneous assembly of empty versus
  functional single-stranded rna viruses.
\newblock \emph{Phys. Rev. E} 78:051915.

\bibitem[{Endres and Zlotnick(2002)}]{Endres2002}
Endres, D., and A.~Zlotnick. 2002.
\newblock Model-based analysis of assembly kinetics for virus capsids or other
  spherical polymers.
\newblock \emph{Biophys. J.} 83:1217--1230.

\bibitem[{Hagan(2009{\natexlab{a}})}]{Hagan2009b}
Hagan, M.~F. 2009{\natexlab{a}}.
\newblock Understanding the concentration dependence of viral capsid assembly
  kinetics - the origin of the lag time and identifying the critical nucleus
  size.
\newblock \emph{Biophys. J.} submitted.

\bibitem[{Hagan(2008)}]{Hagan2008}
Hagan, M.~F. 2008.
\newblock Controlling viral capsid assembly with templating.
\newblock \emph{Phys. Rev. E} 77:051904.

\bibitem[{Hagan(2009{\natexlab{b}})}]{Hagan2009}
Hagan, M.~F. 2009{\natexlab{b}}.
\newblock A theory for viral capsid assembly around electrostatic cores.
\newblock \emph{J. Chem. Phys.} 130:114902.

\bibitem[{D'Souza and Summers(2005)}]{D'Souza2005}
D'Souza, V., and M.~F. Summers. 2005.
\newblock How retroviruses select their genomes.
\newblock \emph{Nat. Rev. Microbiol.} 3:643--655.

\bibitem[{Jouvenet et~al.(2009)Jouvenet, Simon, and Bieniasz}]{Jouvenet2009}
Jouvenet, N., S.~M. Simon, and P.~D. Bieniasz. 2009.
\newblock Imaging the interaction of hiv-1 genomes and gag during assembly of
  individual viral particles.
\newblock \emph{Proc. Natl. Acad. Sci. U. S. A.} 106:19114--19119.

\bibitem[{Grayson et~al.(2006)Grayson, Evilevitch, Inamdar, Purohit, Gelbart,
  Knobler, and Phillips}]{Grayson2006}
Grayson, P., A.~Evilevitch, M.~M. Inamdar, P.~K. Purohit, W.~M. Gelbart, C.~M.
  Knobler, and R.~Phillips. 2006.
\newblock The effect of genome length on ejection forces in bacteriophage
  lambda.
\newblock \emph{Virology} 348:430--436.

\bibitem[{Jouvenet et~al.(2008)Jouvenet, Bieniasz, and Simon}]{Jouvenet2008}
Jouvenet, N., P.~D. Bieniasz, and S.~M. Simon. 2008.
\newblock Imaging the biogenesis of individual hiv-1 virions in live cells.
\newblock \emph{Nature} 454:236--240.

\bibitem[{Hicks and Henley(2006)}]{Hicks2006}
Hicks, S.~D., and C.~L. Henley. 2006.
\newblock Irreversible growth model for virus capsid assembly.
\newblock \emph{Phys. Rev. E} 74:031912.

\bibitem[{Nguyen et~al.(2007)Nguyen, Reddy, and Brooks}]{Nguyen2007}
Nguyen, H.~D., V.~S. Reddy, and C.~L. Brooks. 2007.
\newblock Deciphering the kinetic mechanism of spontaneous self-assembly of
  icosahedral capsids.
\newblock \emph{Nano Lett.} 7:338--344.

\bibitem[{Wilber et~al.(2007)Wilber, Doye, Louis, Noya, Miller, and
  Wong}]{Wilber2007}
Wilber, A.~W., J.~P.~K. Doye, A.~A. Louis, E.~G. Noya, M.~A. Miller, and
  P.~Wong. 2007.
\newblock Reversible self-assembly of patchy particles into monodisperse
  icosahedral clusters.
\newblock \emph{J. Chem. Phys.} 127.

\bibitem[{Nguyen et~al.(2009)Nguyen, Reddy, and Brooks}]{Nguyen2009}
Nguyen, H.~D., V.~S. Reddy, and C.~L. Brooks. 2009.
\newblock Invariant polymorphism in virus capsid assembly.
\newblock \emph{J. Am. Chem. Soc.} 131:2606--14.
\newblock 19199626.

\bibitem[{Carmesin and Kremer(1988)}]{Carmesin1988}
Carmesin, I., and K.~Kremer. 1988.
\newblock The bond fluctuation method: a new effective algorithm for the
  dynamics of polymers in all spatial dimensions.
\newblock \emph{Macromolecules} 21:2819--2823.

\bibitem[{Shaffer(1994)}]{Shaffer1994}
Shaffer, J.~S. 1994.
\newblock Effects of chain topology on polymer dynamics: Bulk melts.
\newblock \emph{J. Chem. Phys.} 101:4205--4213.

\bibitem[{Rosenbluth and Rosenbluth(1955)}]{Rosenbluth1955}
Rosenbluth, M.~N., and A.~W. Rosenbluth. 1955.
\newblock Monte-carlo calculation of the average extension of molecular chains.
\newblock \emph{J. Chem. Phys.} 23:356--359.

\bibitem[{Elrad and Hagan(2008)}]{Elrad2008}
Elrad, O.~M., and M.~F. Hagan. 2008.
\newblock \emph{Nano Lett.} 8:3850--3857.

\bibitem[{Sakaue and Raphael(2006)}]{Sakaue2006}
Sakaue, T., and E.~Raphael. 2006.
\newblock Polymer chains in confined spaces and flow-injection problems: Some
  remarks.
\newblock \emph{Macromolecules} 39:2621--2628.

\bibitem[{Whitelam and Geissler(2007)}]{Whitelam2007}
Whitelam, S., and L.~Geissler, Phillip. 2007.
\newblock Avoiding unphysical kinetic traps in monte carlo simulations of
  strongly attractive particles.
\newblock \emph{J. Chem. Phys.} 127:154101.

\bibitem[{Natarajan et~al.(2005)Natarajan, Lander, Shepherd, Reddy, Brooks, and
  Johnson}]{Natarajan2005}
Natarajan, P., G.~C. Lander, C.~M. Shepherd, V.~S. Reddy, C.~L. Brooks, and
  J.~E. Johnson. 2005.
\newblock Exploring icosahedral virus structures with viper.
\newblock \emph{Nat. Rev. Microbiol.} 3:809--817.

\end{thebibliography}

\clearpage

\end{document}